\documentclass{aa}
\usepackage{txfonts}
\usepackage{graphicx}
\usepackage{float}
\usepackage{natbib}
\bibpunct{(}{)}{;}{a}{}{,}

\newcommand{\ang}{\mbox{${\rm \AA}$~}}
\newcommand{\Teff}{\mbox{$T_{\rm eff}$}}
\newcommand{\Lg}{\mbox{$\log\,g$}}

\newcommand{\Msol}{\mbox{M$_{\odot}$}}

\begin{document}

\title{Spectral analyses of DO white dwarfs and PG 1159 stars from the
  Sloan Digital Sky Survey} \titlerunning{Spectral analyses of DO
  white dwarfs and PG 1159 stars} \author{S.~D. H\"ugelmeyer\inst{1}
  \and S. Dreizler\inst{1} \and K. Werner\inst{2} \and
  J.~Krzesi\'nski\inst{3,4} \and A. Nitta\inst{3} \and
  S.~J. Kleinman\inst{3}}

\institute{Institut f\"ur Astrophysik, Universit\"at G\"ottingen,
  Friedrich-Hund-Platz 1, 37077 G\"ottingen, Germany \and Institut
  f\"ur Astronomie und Astrophysik, Universit\"at T\"ubingen, Sand 1,
  72076 T\"ubingen, Germany \and New Mexico State University, Apache
  Point Observatory, 2001 Apache Point Road, P.O. Box 59, Sunspot, NM
  88349, USA \and Mt. Suhora Observatory, Cracow Pedagogical
  University, ul. Podchorazych 2, 30-084 Cracow, Poland}

\date{Received 22~April~2005 / Accepted 04~July~2005}
 
\abstract{We present a model atmosphere analysis of ten new DO white
  dwarfs and five new PG 1159 stars discovered in the Sloan Digital Sky
  Survey DR\thanks{DR = Data Release}1, DR2 and DR3. This is a
  significant increase in the number of known DOs and PG 1159 stars. DO
  white dwarfs are situated on the white dwarf cooling sequence from
  the upper hot end (\Teff$\, \approx 120\,000$~K) down to the DB gap
  (\Teff$\, \approx 45\,000$~K). PG 1159 stars on the other hand
  feature effective temperatures which exceed \Teff$\,= 65\,000$~K
  with an upper limit of \Teff$\,= 200\,000$~K and are the proposed
  precursors of DO white dwarfs. Improved statistics are necessary to
  investigate the evolutionary link between these two types of
  stars. From optical SDSS spectra effective temperatures, surface
  gravities and element abundances are determined by means of non-LTE
  model atmospheres.

 }

\maketitle

\section{Introduction}

White dwarfs (WDs) represent the final evolutionary stage for 90\% of
all stars (initial mass ${\rm M}_i < 8 \, {\rm
M}_{\small{\odot}}$). Due to very high mass loss at the hot end of the
Asymptotic Giant Branch (AGB), these objects lose part of their
envelope forming a planetary nebula in their subsequent evolution. The
remaining core of the star rapidly evolves toward high effective
temperatures (\mbox{\Teff $\,> 100\,000$~K}). When H- and He-shell
burning ceases, the star enters the WD cooling sequence. The evolution
of these post-AGB objects is separated into a H-rich and H-deficient
sequence, the latter one occurs with a commonness of 20\%. Following
the first spectral classification of \citet{2004A&A...417.1093K}, we
present a spectral analysis of two types of hot He-rich objects from
the Sloan Digital Sky Survey (SDSS).

\subsection{PG 1159 stars}

PG 1159 stars are transition objects between the hottest post-AGB and
the WD phase. The prototype of this spectroscopic class, PG 1159-035,
was found in the Palomar Green (PG) survey
\citep{1986ApJS...61..305G}. It shows a spectrum without detectable
H-absorption lines. In fact, it is dominated by \ion{He}{ii} and
highly ionised carbon and oxygen lines. PG 1159 stars are characterised
by a broad absorption trough around 4760~\ang composed by \ion{He}{ii}
4686~\ang and several \ion{C}{iv} lines, suggesting high effective
temperatures. Spectral analyses reveal \Teff$\,=65\,000$--$200\,000$~K
and gravities of \Lg \,= 5.5--8.0
\citep{1991A&A...244..437W,1994A&A...286..463D,1996aeu..conf..229W}.
From the 28 currently (prior to SDSS) known PG 1159 stars, ten are
low-gravity \citep[subtype lgE, ][]{1992LNP...401..273W} stars in the
region of hot central stars of planetary nebulae (CSPNe) while the
others are more compact objects with surface gravities of WDs (subtype
A or E). The majority of the PG 1159 stars was discovered in large
surveys \citep[Palomar Green, Hamburg Schmidt (HS),
][]{1995A&AS..111..195H}. The most recent and only discovery
\citep{2004A&A...424..657W} within the last 10 years was an object
from the Hamburg ESO (HE) survey
\citep{1996A&AS..115..227W}. Currently, the SDSS
\citep{2000AJ....120.1579Y} offers a new opportunity to increase the
number of known PG 1159 stars.

PG 1159-035 (= GW Vir) also defines a new class of variable
stars. \citet{1979wdvd.coll..377M} discovered low-amplitude non-radial
g-mode pulsations. About one third of the PG 1159 stars show this
variability driven by cyclic ionisation of carbon and oxygen
\citep{1986HiA.....7..229C,1987fbs..conf..309S}.  Analyses of HST
spectra indicate that stars with high carbon and oxygen abundance are
more likely to pulsate \citep{1998A&A...334..618D} which is
corroborated by theoretical calculations of
\citet{2004ApJ...610..436Q}.

The PG 1159 region in the Hertzsprung-Russell-Diagram (HRD) overlaps
with that of DO white dwarfs. Therefore it is assumed that
gravitational settling of the heavier elements in the atmosphere of
the PG 1159 stars leads to the transition towards DO white dwarfs.

\subsection{DO white dwarfs}

White dwarfs can be separated into two distinct spectroscopic classes,
DA and non-DA white dwarfs. The former ones show a pure hydrogen
spectrum and can be found on the entire WD cooling sequence. The
latter ones fall into three subclasses. DO dwarfs with
\mbox{$45\,000$~K\,$<$ \Teff\,$< 120\,000$~K}, DB stars
(\mbox{$11\,000$~K\,$<$ \Teff\,$< 30\,000$~K}) and DCs
(\mbox{\Teff\,$< 11\,000$~K}). The spectroscopic appearance of each
class is determined by the ionisation balance of \ion{He}{i} and
\ion{He}{ii}. DO white dwarfs show a pure \ion{He}{ii} spectrum at the
hot end and a mixed \ion{He}{i/ii} spectrum at the cool end. The
transition to the cooler DB dwarfs, characterised by pure \ion{He}{i}
spectra, is interrupted by the so-called \mbox{``DB gap''}
\citep{1986ApJ...309..241L}. In the HRD region of stars with
\mbox{$30\,000$~K \,$<$ \Teff \, $< 45\,000$~K} no objects with
H-deficient atmospheres have been observed to date. This phenomenon is
a fundamental problem in the understanding of WD spectral evolution.

As in the case of PG 1159 stars, the most recent discovery of a DO
white dwarf is due to the HE survey \citep{2004A&A...424..657W}, while
the PG and HS surveys contributed the majority in time steps of
decades with 19 DOs known prior to the SDSS.

\section{SDSS Observations}

The SDSS is a photometric and spectroscopic survey covering ~7000
square degrees of the sky around the northern Galactic cap
\citep{2000AJ....120.1579Y}.  The survey first images the sky in five
passbands and uses these data to select interesting targets for
spectroscopic follow-up. The survey's main goal is to study the large
scale structure of the universe, therefore only a small fraction of
the observed stars are targeted for spectroscopy.  For
spectrophotometric calibration purposes, however, certain
``HOT$\_$STD'', or hot standard stars, are specifically
targeted. These objects meet the following photometric criteria:
\mbox{$g>14$, $g_o<19$} (where subscript~$o$ denotes a dereddened
magnitude), \mbox{$-1.5<(u-g)_o<0.0$}, and \mbox{$-1.5<(g-r)_o <0.0$.}
Since the number of objects which meet these criteria is relatively
small, such objects are targeted nearly to completion.  All but six
of the DOs and PG 1159 stars in this paper were observed by the SDSS as
hot standard stars and reported first by \citet{2004A&A...417.1093K}
while the remaining six come from HOT$\_$STD spectra included in the
Second and Third Data Release of the SDSS
\citep{2004AJ....128..502A,2005AJ....129.1755A} not
analysed in the \citet{2004A&A...417.1093K} work. A detailed
description of the SDSS spectrographs and spectral data can be found
in \citet{2002AJ....123..485S} and
\citet{2003AJ....126.2081A,2004AJ....128..502A,2005AJ....129.1755A}. In
short, the SDSS spectral data cover a wavelength range from 3800 to
9200\,~\AA\ with R$\,\sim\,$1800.  They are flux calibrated to about
10\% and have an average signal-to-noise ratio of $\sim\,$4 at
$g=20.2$.

\section{Spectral Analysis}

In order to analyse the DO and PG 1159 spectra, we calculated
homogeneous, plane-parallel non-LTE model atmospheres with a code
based on the Accelerated Lambda Iteration \citep[][ and references
therein]{2003ASPC..288...31W}. For these types of stars it is
necessary to account for non-LTE effects as shown for DO white dwarfs
by \citet{1996A&A...314..217D} and for PG 1159 stars by
\citet{1991A&A...244..437W}. For comparison of synthetic and observed
spectra, the latter ones are normalised using a third order polynomial
fit through the continuum points, which are determined using the
normalised theoretical spectra. Lineshifts due to radial velocities
are taken into account by means of cross-correlation.  This comparison
procedure is performed by an \verb|IDL| code routine in order to
guarantee consistent results. For the DO white dwarfs we used a
$\chi^2$-statistic to derive best-fit models and 1-$\sigma$ errors
following \citet{1986ApJ...305..740Z}. Compared to our preliminary
analyses \citep{pwd04...sdh} with best-fit models selected by eye, we
find differences especially for the hot DO white dwarfs. The sparsely
populated model grid for the PG 1159 stars does not allow a reasonable
application of $\chi^2$-statistics. We therefore have to rely on
best-fit models selected by eye, guided by the variance of model $-$
observation. Error estimates are obtained from a global analysis of
the goodness of fit for neighbouring parameter sets.

\subsection{Spectral analysis of PG 1159 stars}

\begin{figure*}
  \centering \includegraphics[bb=73 365 545
  709,width=18cm,clip]{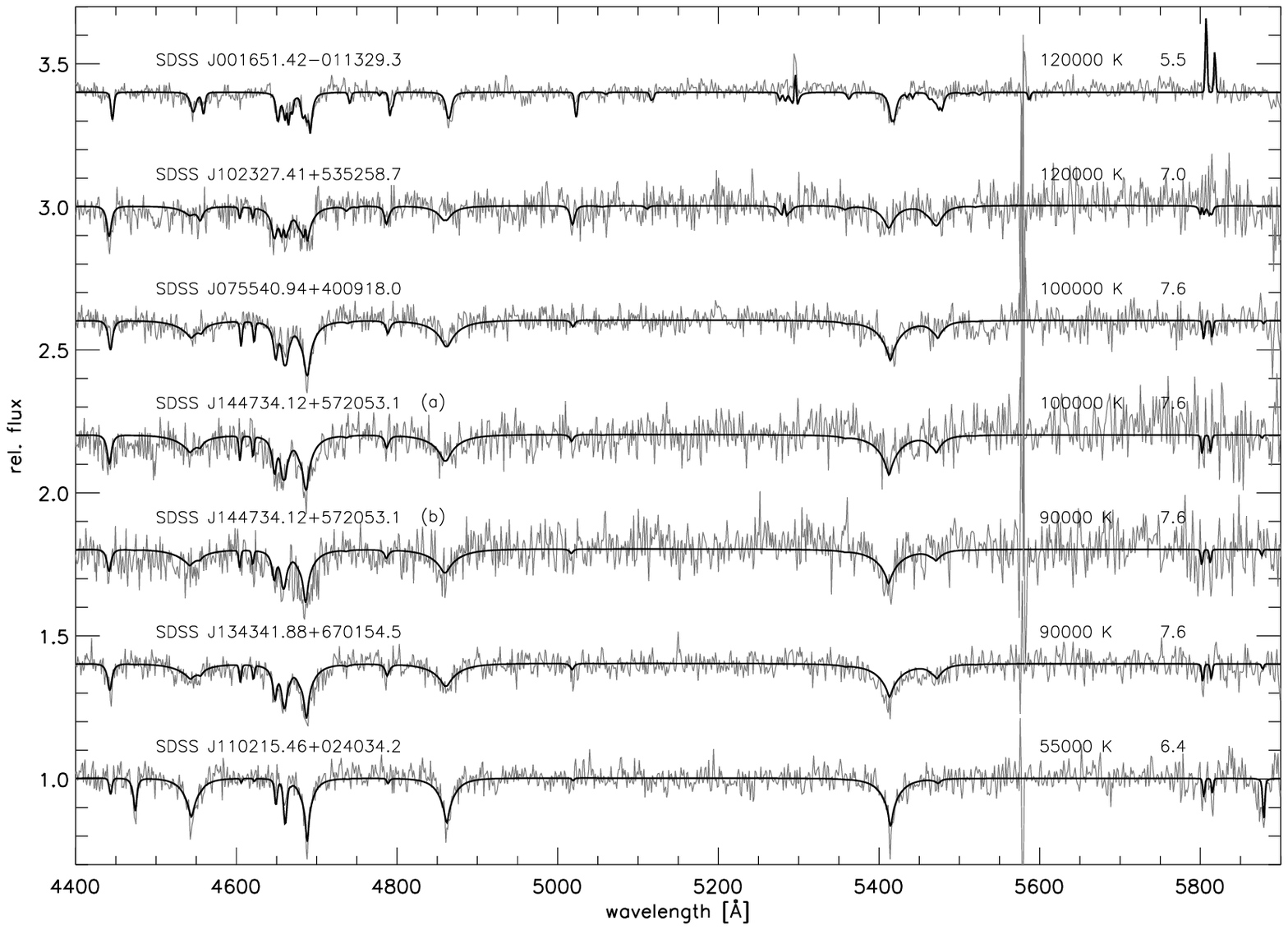}
  \caption{Normalised optical spectra (grey lines) of PG 1159 stars and
    the sdO star (bottom-most spectrum) and model atmospheres (black
    lines), ordered by decreasing effective temperature. Object names
    are printed on the left, effective temperatures and logarithmic
    surface gravities on the right. SDSS J144734.12 +572053.1 was
    observed twice on different plates and slightly different
    temperatures are derived, however, within our error estimate.}
  \label{fig:pgs}
\end{figure*}

For the PG 1159 stars we calculated atmospheres using detailed H-He-C-O
model atoms (Fig.~\ref{fig:pgs}). The model grid ranges from
\mbox{\Teff\,$= 55\,000$--$150\,000$~K} and \mbox{\Lg\,$=
5.5$--$7.6$}. A complete coverage of this parameter space is not
available due to high computational time for all model
atmospheres. The abundances are fixed to values He/H\,=\,100 and
C/He\,=\,0.01, 0.03, 0.05, 0.1, 0.3, 0.6 by number. While oxygen can only be
determined in the hottest PG 1159 star, the best-fit models for the
rest of the PG 1159 candidates were calculated with an oxygen abundance
following the typical PG 1159 abundance-scaling ratio
O/C$\,\approx\,$C/He. However, variations in the oxygen abundance do
not show a significant effect on the other stellar parameters.

\subsection{Spectral analysis of DO white dwarfs}

\begin{figure*}
  \centering \includegraphics[bb=75 365 545
  709,width=18cm,clip]{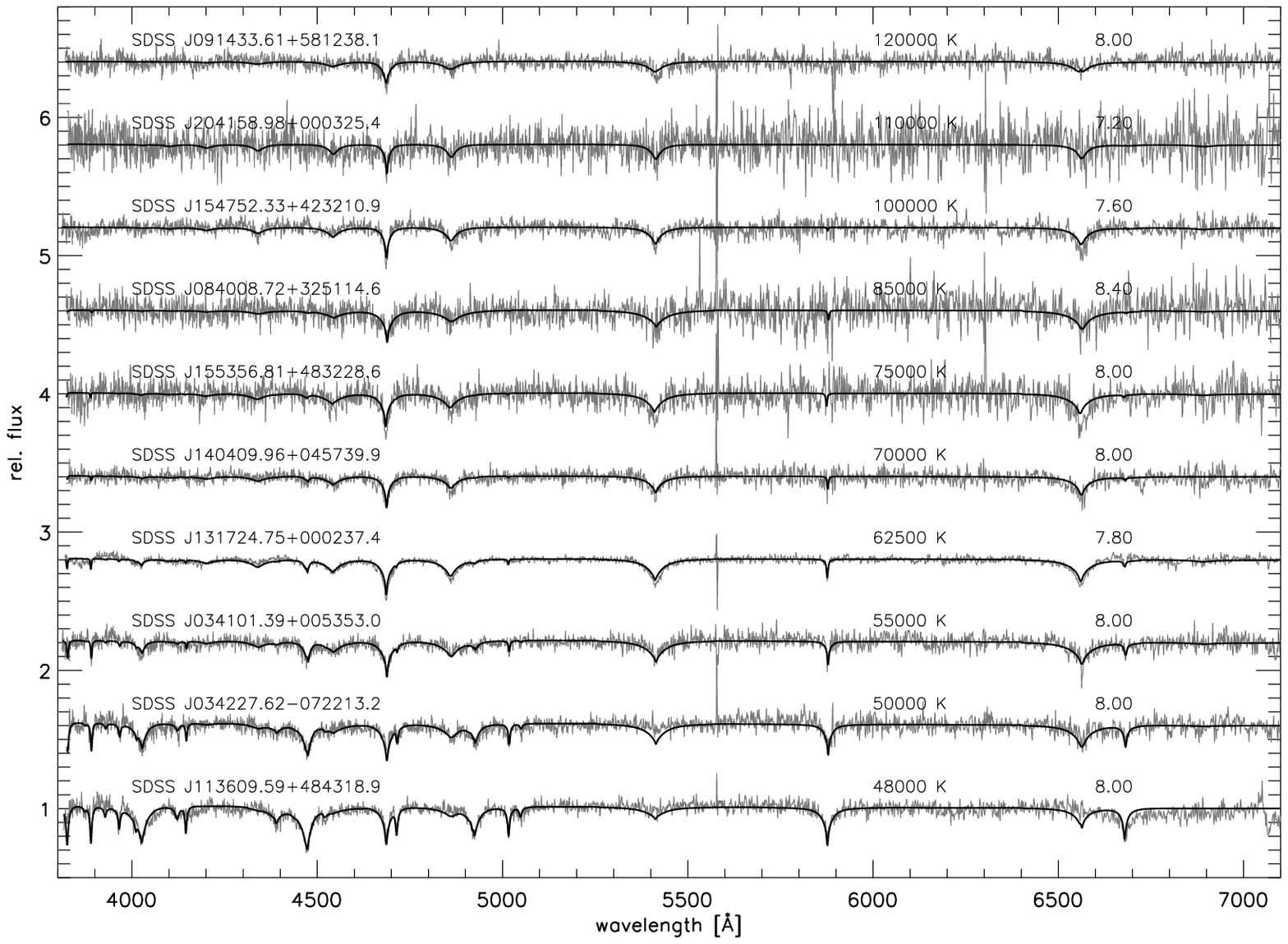}
  \caption{Normalised optical spectra (grey lines) of DO white dwarfs
  and model atmospheres (black lines), ordered by decreasing effective
  temperature. Object names are printed on the left, effective
  temperatures and logarithmic surface gravities on the right. The
  spectra of the hottest stars are dominated by \ion{He}{ii}
  absorption lines, while increasing line strengths of \ion{He}{i} are
  observed with decreasing temperature.}
  \label{fig:dos}
\end{figure*}

Detailed H-He atomic models \citep{1996A&A...314..217D} are used to
calculate the model atmospheres (Fig.~\ref{fig:dos}). The model grid
ranges from \mbox{\Teff\,$= 42\,500$--$120\,000$~K} in steps of
$2\,500$~K. The \Lg\, ranges from 7.0 to 8.4 in intervals of 0.2. The
helium abundance is fixed to He/H\,=\,99.

\section{Results and Discussion}

Starting from spectral classification of hydrogen-deficient white
dwarfs within DR1 \citep{2004A&A...417.1093K}, we extend this work to
a spectral analysis including similar objects from DR2 and DR3. The
emphasis is placed on PG 1159 stars and DO white dwarfs.

\subsection{PG 1159 stars}

From the seven spectra originally classified as PG 1159 stars, we were
able to confirm five new PG 1159 stars. The results of the spectral
analyses are listed in Table~{\ref{table:1}}.  One of the objects
(\mbox{SDSS J144734.12 +572053.1}) has been observed twice on
different plates. Both spectra are presented here (see
Fig.~\ref{fig:pgs}). The independent analyses of the two spectra
revealed parameters consistent within our error estimates. Despite
strong \ion{C}{iv} lines, one of the objects from our PG 1159 star
candidates (\mbox{SDSS J110215.46 +024034.2}) turned out to be a sdO
star according to its stellar parameters.

Compared to the number of previously known PG 1159 stars, our sample
marks an increase of 18\%.  Following the spectroscopic subtypes
scheme of \citet{1992LNP...401..273W}, \mbox{SDSS J001651.42
$-$011329.3} was classified as lgE type, \mbox{SDSS J102327.41
+535258.7} as an E and the remaining three as A types The latter ones
show very similar temperatures and gravities. Regarding the carbon
abundance, the PG 1159 stars clearly fall into two groups: one with
carbon abundances C/He\,$\approx$\,0.2\,--\,0.3 (by number) and the
other one with C/He\,$\approx$\,0.03\,--\,0.05. This separation
underlines earlier results of PG 1159 stars gained from HST spectra
\citep{1998A&A...334..618D}. These previous analyses indicated a
correlation between high carbon abundances and presence of pulsations,
which is also corroborated by the theoretical investigations of
\citet{2004ApJ...610..436Q}. Using the above mentioned correlation, we
predict that \mbox{SDSS J001651.42 $-$011329.3} and \mbox{SDSS
J102327.41 +535258.7} are pulsators.

While carbon and helium abundances can be derived from optical spectra
of PG 1159 stars, the third most abundant element, oxygen, cannot be
analysed in type A PG 1159 stars due to the lack of sufficiently strong
spectral lines. \ion{O}{vi} lines, present in hotter PG 1159 stars, are
not excited while transitions of lower oxygen ionisation stages are
weak in the optical range. Therefore, only the hottest object of the
sample, \mbox{SDSS J001651.42$-$011329.3}, allows the determination of
an oxygen abundance (O/He\,=\,0.04). For the others, additional UV
spectra would be required.

\begin{figure}
  \centering \includegraphics[bb=129 368 863
  1092,width=8.8cm,clip]{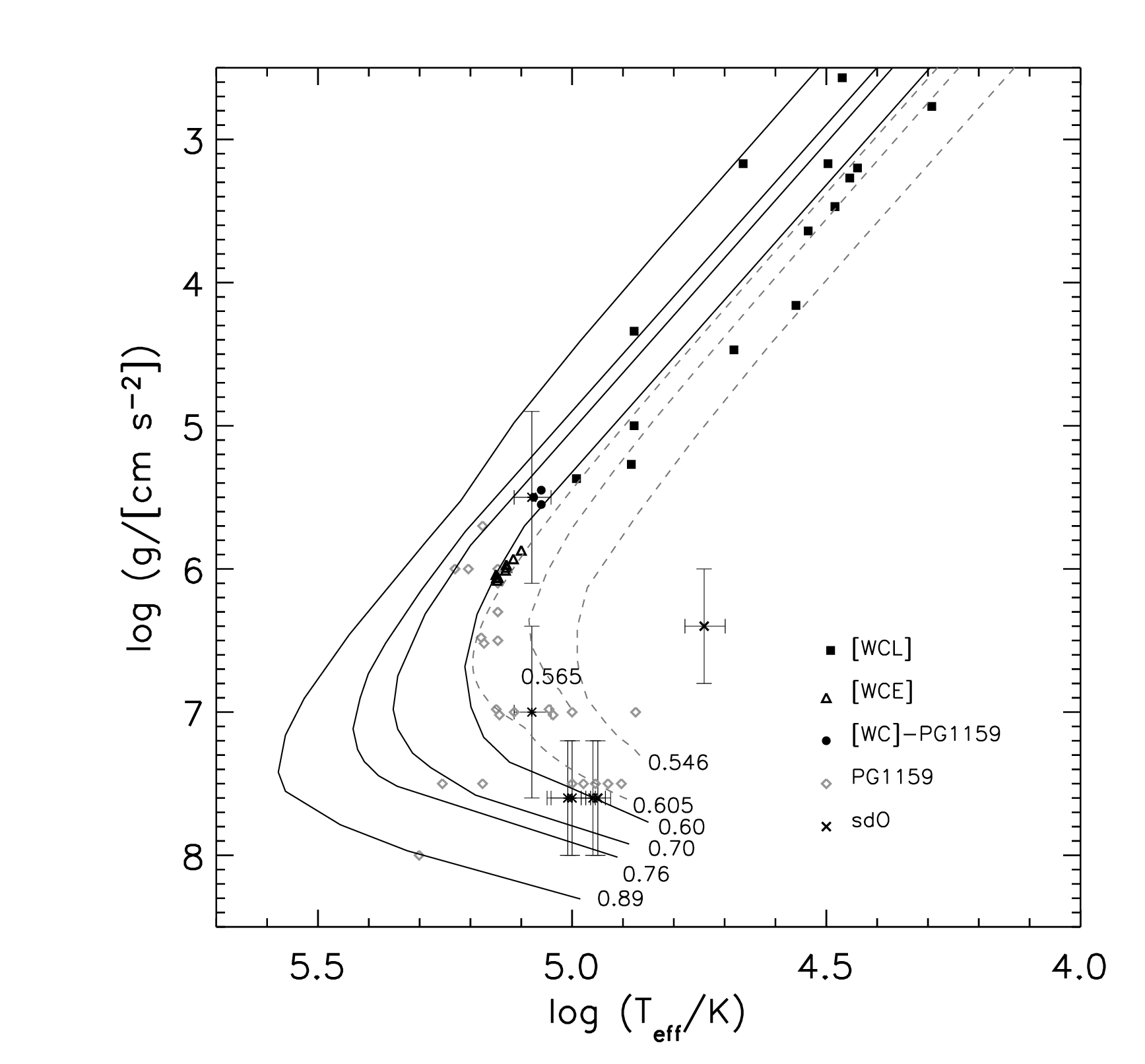}
  \caption{Positions of the new PG 1159 stars (with error bars)
    compared to evolutionary tracks from \citet{1995A&A...299..755B},
    \citet{1983ApJ...272..708S} (dashed lines) and
    \citet*{1986ApJ...307..659W}. Labels: mass in
    M$_{\small{\odot}}$.}
  \label{fig:trackpg}
\end{figure}

Finally, we compare the positions of the new PG 1159 stars with
evolutionary tracks as well as with positions of previously known
PG 1159 stars and related objects (Fig.~\ref{fig:trackpg}). From our
results \mbox{SDSS J001651.42 $-$011329.3} marks the transition from
[WC]-PG 1159 to PG 1159 stars. From its position in the
\mbox{\Teff--\Lg} \, diagram (Fig.~\ref{fig:trackpg}), the sdO star
did clearly not have an AGB history but did rather evolve directly
from the Horizontal Branch.

\begin{table}
  \caption{Atmospheric parameters of PG 1159 stars and the sdO star
    (last entry). The C/He abundance ratio is given by number.}
  \label{table:1}
  \centering
  \begin{tabular}{l r @{\,$\pm$\,} l r @{\,$\pm$\,} l c}
    \hline\hline\noalign{\smallskip} Name & \multicolumn{2}{c}{T$_{\rm
    eff}$} & \multicolumn{2}{c}{$\log g$} & C/He\\ &
    \multicolumn{2}{c}{[kK]} & \multicolumn{2}{c}{[cgs]} & \\ \hline
    SDSS J001651.42$-$011329.3 & 120&10 & 5.5&0.6 & 0.20 \\ SDSS
    J102327.41+535258.7 & 120&10 & 7.0&0.6 & 0.30 \\ SDSS
    J075540.94+400918.0 & 100&10 & 7.6&0.4 & 0.03 \\ SDSS
    J144734.12+572053.1 (a) & 100&10 & 7.6&0.4 & 0.05 \\ SDSS
    J144734.12+572053.1 (b) & 90&\, 5 & 7.6&0.4 & 0.03 \\ SDSS
    J134341.88+670154.5 & 90&\, 5 & 7.6&0.4 & 0.05 \\ \hline SDSS
    J110215.46+024034.2 & 55&\, 5 & 6.4&0.4 & 0.01 \\ \hline
  \end{tabular}
\end{table}

\subsection{DO white dwarfs}

The stellar parameters are presented in Table~\ref{table:2}.  From the
ten objects -- an increase of 50\% of known DO white dwarfs -- one was
previously known from the Hamburg ESO Survey \citep[SDSS
J131724.75+000237.4\,=\,HE1314+0018,
][]{2004A&A...424..657W}. Comparing our results with theirs, we find
similar stellar parameters. The UVES/VLT spectrum of
\mbox{HE1314+0018} analysed by \citet{2004A&A...424..657W}, however,
allowed a much more detailed comparison between synthetic spectra and
observation, showing discrepancies for all available models. In our
low resolution spectrum, this finding can be confirmed, excluding data
reduction problems as a solution for the systematic deviation between
models and observation. This phenomenon has previously been observed
in DO-like stars with signatures of very highly excited C, N, O, and
Ne lines \citep{1995A&A...293L..75W,1995A&A...303L..53D}, however, no
such spectral lines are present in \mbox{HE1314+0018}.

Two stars, \mbox{SDSS J131724.75+000237.4} and SDSS
J140409.96+045739.9, provide detectable \ion{C}{iv} lines. While the
carbon abundance of the former one was already determined by
\citet{2004A&A...424..657W} to C/He\,=\,0.001, the latter one features
a carbon abundance of C/He\,=\,0.01 \citep{pwd04...sdh}. The spectra
of the remaining objects do not allow to derive metal abundances due
to low signal-to-noise ratios.

\begin{figure}
  \centering \includegraphics[bb=111 372 845
  1092,width=8.8cm,clip]{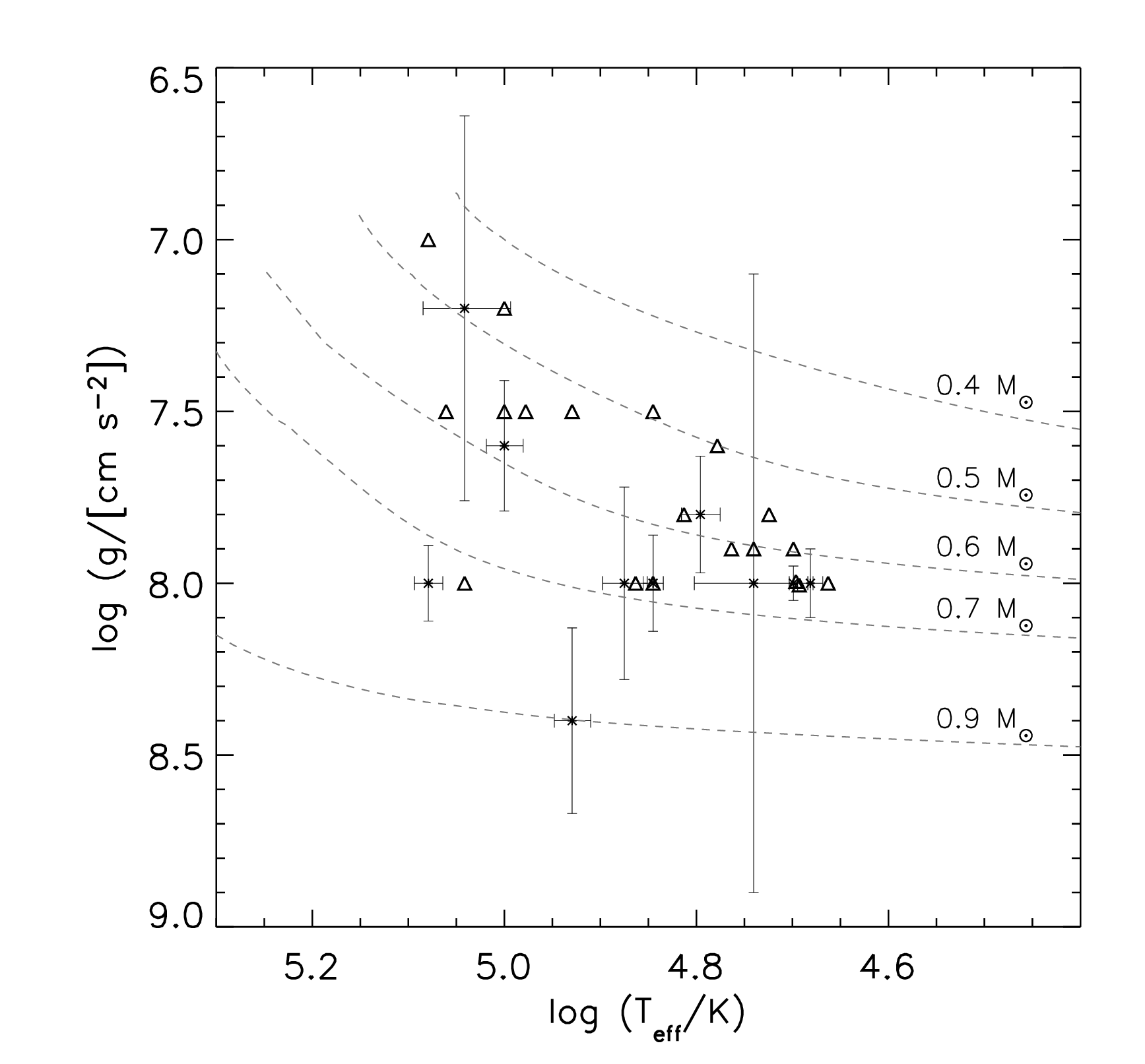}
  \caption{Positions of DO white dwarfs compared with evolutionary
  tracks from \citet{1995LNP...443...41W}. The triangles represent the
  19 hitherto known DOs
  \citep[see][]{1996A&A...314..217D,1997fbs..conf..303D,2004A&A...424..657W}.}
  \label{fig:trackdo}
\end{figure}

\begin{figure}
  \centering \includegraphics[bb=80 366 543
  706,width=8.8cm,clip]{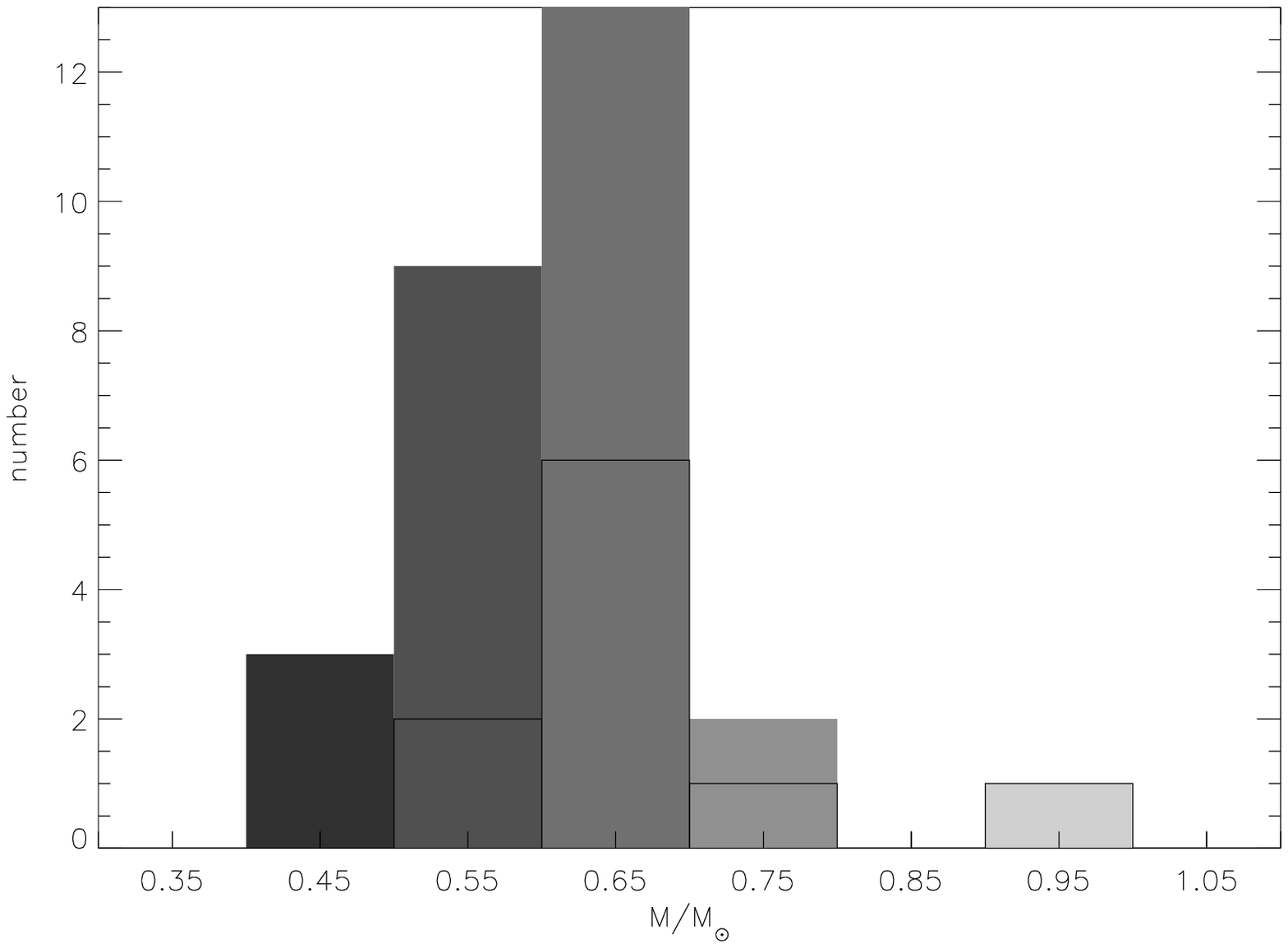}
  \caption{Mass distribution (bin\,=\,0.1\,\Msol) of all known DO
    white dwarfs. The region below the back line represents masses of
    stars from this work while the area above the line shows results
    from \citet{1996A&A...314..217D} and \citet{1997fbs..conf..303D}.}
  \label{fig:massdis}
\end{figure}

Stellar masses are derived from our atmospheric parameters using an
interpolation program for evolutionary tracks
\citep{1995LNP...443...41W} written by D. Koester. The results are
displayed in Fig.~\ref{fig:trackdo}, Fig.~\ref{fig:massdis} and
Table~\ref{table:2}. We identified the most massive DO white dwarf
(0.9 \Msol) known so far. In general, the mass distribution of SDSS DO
white dwarfs seems to be slightly shifted towards higher masses. This
suggests a consistent reanalysis of the other DO stars with our
current model grid and $\chi^2$-technique. Using the same program by
D. Koester, we also derived cooling ages for the DO white dwarfs in
the range of 145\,000 to 2.500\,000 years. A complete statistical
analysis of DO white dwarfs is planned for presentation after the last
SDSS data release.

\begin{table}
  \caption{Stellar parameters of DO white dwarfs.}
  \label{table:2}
  \centering
  \begin{tabular}{l r @{\,$\pm$\,} l r @{\,$\pm$\,} l c}
    \hline\noalign{\smallskip} Name & \multicolumn{2}{c}{T$_{\rm
    eff}$} & \multicolumn{2}{c}{$\log g$} & M \\ &
    \multicolumn{2}{c}{[kK]} & \multicolumn{2}{c}{[cgs]} & [\Msol]\\
    \hline SDSS J091433.61+581238.1 & 120.0&\, 4.1 & 8.00&0.11 & 0.75
    \\ SDSS J204158.98+000325.4 & 110.0&11.5 & 7.20&0.56 & 0.60 \\
    SDSS J154752.33+423210.9 & 100.0&\, 4.4 & 7.60&0.19 & 0.59 \\ SDSS
    J084008.72+325114.6 & 85.0&\, 3.7 & 8.40&0.27 & 0.90 \\ SDSS
    J155356.81+433228.6 & 75.0&\, 4.0 & 8.00&0.28 & 0.68 \\ SDSS
    J140409.96+045739.9 & 70.0&\, 1.7 & 8.00&0.14 & 0.68 \\ SDSS
    J131724.75+000237.4 & 62.5&\, 2.9 & 7.80&0.17 & 0.58 \\ SDSS
    J034101.39+005353.0 & 55.0&\, 8.4 & 8.00&0.90 & 0.65 \\ SDSS
    J034227.62$-$072213.2 & 50.0&\, 0.5 & 8.00&0.05 & 0.65 \\ SDSS
    J113609.59+484318.9 & 48.0&\, 0.3 & 8.00&0.10 & 0.64 \\ \hline
  \end{tabular}
\end{table}

\section{Acknowledgements}

Funding for the creation and distribution of the SDSS Archive has been
provided by the Alfred P. Sloan Foundation, the Participating
Institutions, the National Aeronautics and Space Administration, the
National Science Foundation, the U.S. Department of Energy, the
Japanese Monbukagakusho, and the Max Planck Society. The SDSS Web site
is http://www.sdss.org/.

The SDSS is managed by the Astrophysical Research Consortium (ARC) for
the Participating Institutions. The Participating Institutions are The
University of Chicago, Fermilab, the Institute for Advanced Study, the
Japan Participation Group, The Johns Hopkins University, the Korean
Scientist Group, Los Alamos National Laboratory, the
Max-Planck-Institute for Astronomy (MPIA), the Max-Planck-Institute
for Astrophysics (MPA), New Mexico State University, University of
Pittsburgh, Princeton University, the United States Naval Observatory,
and the University of Washington.

\end{document}